\documentclass[aps,prb,reprint,superscriptaddress,floatfix]{revtex4-2}

\usepackage{natbib}
\usepackage[dvipdfmx]{graphicx}
\usepackage{bm}
\usepackage{amsmath, amssymb}
\usepackage{upgreek}
\usepackage{color}

\newcommand{\FST}{FeSe$_{1-x}$Te$_{x}$}
\newcommand{\FSS}{FeSe$_{1-x}$S$_{x}$}
\newcommand{\CF}{CaF$_{2}$}
\newcommand{\Ts}{$T_{\mathrm{s}}$}
\newcommand{\Tc}{$T_{\mathrm{c}}$}
\newcommand{\cm}{cm$^{-1}$}
\newcommand{\wn}{$\omega_{\mathrm{p,n}}^2$}
\newcommand{\wb}{$\omega_{\mathrm{p,b}}^2$}
\newcommand{\wratio}{$\omega_{\mathrm{p,n}}^2$ / ($\omega_{\mathrm{p,n}}^2$ + $\omega_{\mathrm{p,b}}^2$)}

\begin{document}


\title{Evolution of charge dynamics in \FST{}: effects of electronic correlations and nematicity}

\author{M.~Nakajima}
\email[]{nakajima@phys.sci.osaka-u.ac.jp}

\author{K.~Yanase}
\affiliation{Department of Physics, Osaka University, Osaka 560-0043, Japan}

\author{M.~Kawai}
\author{D.~Asami}
\author{T.~Ishikawa}
\author{F.~Nabeshima}
\affiliation{Department of Basic Science, The University of Tokyo, Tokyo 153-8902, Japan}

\author{Y.~Imai}
\affiliation{Department of Physics, Graduate School of Science, Tohoku University, Sendai 980-8578, Japan}

\author{A.~Maeda}
\affiliation{Department of Basic Science, The University of Tokyo, Tokyo 153-8902, Japan}

\author{S.~Tajima}
\affiliation{Department of Physics, Osaka University, Osaka 560-0043, Japan}

\date{\today}

\begin{abstract}
We systematically studied in-plane optical conductivity of \FST{} thin films fabricated on \CF{} substrates for $x$ = 0, 0.1, 0.2, and 0.4. This system shows a large enhancement of superconducting transition temperature \Tc{} at $x \sim$ 0.2 and a gentle decrease in \Tc{} with further increasing $x$. The low-energy optical conductivity spectrum is described by the sum of narrow and broad Drude components, associated with coherent and incoherent charge dynamics, respectively. With increasing Te content, the spectral weight of the narrow Drude component decreases, while the total weight of the two Drude components increases. As a consequence, the fraction of the narrow Drude weight significantly decreases, indicating that Te substitution leads to stronger electronic correlations. Below the nematic transition temperature, the narrow Drude weight decreases with decreasing temperature. This indicates the reduction of the coherent carrier density, resulting from the Fermi-surface modification induced by the development of the orbital order. The reduction of the narrow Drude weight with temperature stopped at $x \sim$ 0.2, corresponding to the disappearance of the nematic transition. Our result suggests that the increase in the coherent carrier density induced by the suppression of the nematic transition gives rise to the enhancement of \Tc{}. The decrease in \Tc{} with further Te substitution likely arises from too strong electronic correlations, which are not favorable for superconductivity.
\end{abstract}

\maketitle

\section{Introduction}

Iron-based superconductors (FeSCs) are characterized by a rich phase diagram. To deeply understand the underlying physics, it is indispensable to extract key control parameters governing phase diagrams and to elucidate how an electronic state evolves with these parameters. In many FeSCs, superconductivity is induced by suppressing a phase in which four-fold rotational symmetry is broken, which suggests a close relationship between an anisotropic electronic state and superconductivity~\cite{Fernandes2014}.

FeSe exhibits a tetragonal-to-orthorhombic structural phase transition at \Ts{} $\sim$ 90 K without accompanying a magnetic phase transition~\cite{McQueen2009}, followed by a superconducting transition at \Tc{} $\sim$ 8 K~\cite{Hsu2008}. Here, the effect of static magnetism can be neglected from discussion. In the orthorhombic phase, the splitting of the $d_{xz}$ and $d_{yz}$ bands, which are equivalent in the high-temperature tetragonal phase, was observed, indicative of the presence of orbital ordering~\cite{Shimojima2014,Nakayama2014,Watson2015}. The structural transition is electronic in origin and thus is often referred to as a nematic transition.

The nematic transition in FeSe can be tuned by chemical substitution~\cite{Mizuguchi2009,Imai2015}, hydrostatic pressure~\cite{Medvedev2009,Sun2016}, and in-plane biaxial strain~\cite{Nabeshima2018,Nakajima2021}. To study the influence of electronic nematicity on superconductivity, isovalent S substitution and physical pressure have been employed so far, but this is not straightforward to understand the electronic state because magnetism is simultaneously involved~\cite{Kothapalli2016,Nabeshima2018b,Yi2020,Nabeshima2021}. (Note that the magnetic phase adjoins the nematic phase for thin films of \FSS{}~\cite{Nabeshima2018b} but not for single crystals~\cite{Yi2020}.) On the other hand, no magnetism has been observed for Te substitution~\cite{Imai2015,Imai2017,Terao2019,Mukasa2021}. For the \FST{} films on \CF{} substrates, \Tc{} is largely enhanced at $x \sim$ 0.2 and takes a maximum value of 23 K [Fig.~1(a)]. It was suggested that the nematic transition disappears at this composition~\cite{Imai2017}, which was indeed confirmed by angle-resolved photoemission spectroscopy (ARPES) measurements~\cite{Nakayama2021}. \FST{} is thus a suitable system to investigate the relationship between electronic nematicity and superconductivity.

Another important issue in FeSCs is electronic correlation. According to the theoretical calculations, a superconducting phase can emerge in the vicinity of a strongly correlated electron regime centered at a Mott insulating phase with the $d^5$ configuration~\cite{Misawa2012,Medici2014}. Combined with their multiorbital nature, electronic correlations in iron-based compounds are significantly orbital selective~\cite{Yin2011,Misawa2012,Medici2014}. In particular, electrons with the $d_{xy}$ orbital character are strongly correlated. Hund's coupling plays a role in an enhancement of the orbital differentiation. Among iron-based compounds, the correlation strength and the orbital differentiation are largest in iron chalcogenides~\cite{Yi2017}. It has been found that Te-rich \FST{} shows an incoherent-to-coherent crossover in the electronic structure as Te content as well as temperature is decreased~\cite{Yi2015,Liu2015,Otsuka2019}. How electronic correlations in FeSe evolve with Te substitution is an issue to be addressed.

In the present study, we performed optical spectroscopy measurements on \FST{} thin films on \CF{} substrates with $x$ = 0, 0.1, 0.2, and 0.4. Optical spectroscopy is a useful bulk-sensitive probe to investigate charge dynamics. As in the case for other FeSCs, the low-energy optical conductivity spectrum of \FST{} can be decomposed into a narrow and a broad Drude component. The narrow Drude component is overdamped at high temperatures, resulting in a structureless flat spectrum in the low-energy region characterized by highly incoherent charge dynamics. The weight of the narrow Drude component decreases with increasing $x$, while the sum of the weights of the two Drude components increases. This gives rise to a decrease in the fraction of the narrow Drude weight, indicating that electronic correlations become stronger with Te substitution. With decreasing temperature across \Ts{}, a decrease in the narrow Drude weight was observed. This behavior is suppressed with substituting Te for Se, leading to a disappearance of \Ts{} at $x \sim$ 0.2. Since the narrow Drude weight corresponds to coherent carrier density, the suppression of the nematic transition increases the coherent carrier density, which is likely related with the enhancement of \Tc{} at $x \sim$ 0.2.

\section{Experimental}

\begin{figure}
\includegraphics[width=6.7cm]{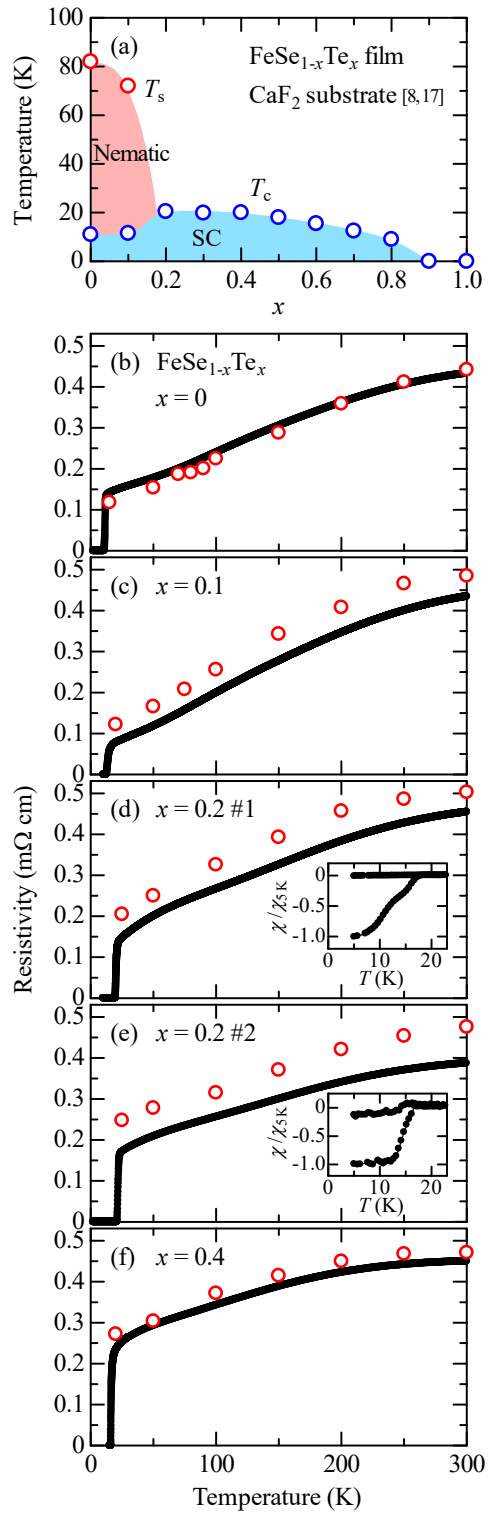}%
\caption{\label{} (a) Phase diagram of the \FST{} thin films on \CF{} substrates as a function of Te content $x$~\cite{Imai2015,Imai2017}. Temperature dependence of resistivity for the \FST{} thin films for (b) $x=0$, (c) 0.1, (d) 0.2 \#1, (e) 0.2 \#2, and (f) 0.4. Red circles indicate the resistivity values calculated from the dc conductivity obtained by the Drude-Lorentz analysis. Insets of (d) and (e) show the temperature dependence of magnetic susceptibility normalized to the zero-field-cooled value at 5 K.}
\end{figure}

Thin films of \FST{} ($x$ = 0, 0.1, 0.2, and 0.4) were fabricated on \CF{} substrates by a pulsed laser deposition method~\cite{Imai2015}. Two samples were prepared for $x$ = 0.2 (\#1 and \#2), at which the highest \Tc{} is observed for this system. The thicknesses of the grown films were 205, 210, 150, 170, and 190 nm for $x$ = 0, 0.1, 0.2 \#1, 0.2 \#2, and 0.4, respectively. \Tc{} was determined from the temperature dependence of resistivity $\rho{}(T)$ [Figs. 1(b)--1(f)]. \Tc{} for $x$ = 0 was 11 K, which is higher than the value for single-crystalline FeSe due to compressive in-plane strain~\cite{Imai2015,Nabeshima2018}. On going from $x$ = 0.1 to 0.2, \Tc{} was abruptly enhanced from 13 K to 20 K (\#1) and 21 K (\#2). The difference of \Tc{} between the two samples for $x$ = 0.2 originates not from a variation of Te content but from different degrees of in-plane strain~\cite{Imai2015,Imai2017}. With further increasing $x$, \Tc{} slightly decreased to 16 K for $x$ = 0.4. Magnetic susceptibility was measured on the two samples for $x$ = 0.2 under a magnetic field of 10 Oe [insets of Figs.~1(d) and 1(e)]. A two-step transition was discernible for \#1, indicating that the sample contains a lower-\Tc{} part. Note that the magnetic susceptibility measurements were performed two and eighteen months after the sample fabrication for \#1 and \#2, respectively, and \Tc{} is reduced due to aging.

Optical reflectivity was measured on the \FST{} films for the energy range of 40--10000 \cm{} using the Fourier-transform infrared spectrometer (Bruker Vertex 80v) at various temperatures ranging from 8 to 300 K. Since the films are oriented along the $c$ axis, in-plane reflectivity spectra were obtained. We extracted optical constants of \FST{} from the reflectivity data by means of a thin-film fitting procedure~\cite{Nakajima2017}. We measured the optical spectrum of \CF{} before the measurement of the thin films and obtained its dielectric function via the Kramers-Kronig transformation. Then, taking into account the reflection from the interface, the thin-film reflectivity spectrum was fitted using a number of Lorentz oscillators. This method does not require the extrapolation of the reflectivity spectrum.

\section{Results and Discussion}

\subsection{Temperature dependence of optical conductivity spectra}

\begin{figure}
\includegraphics[width=7.8cm]{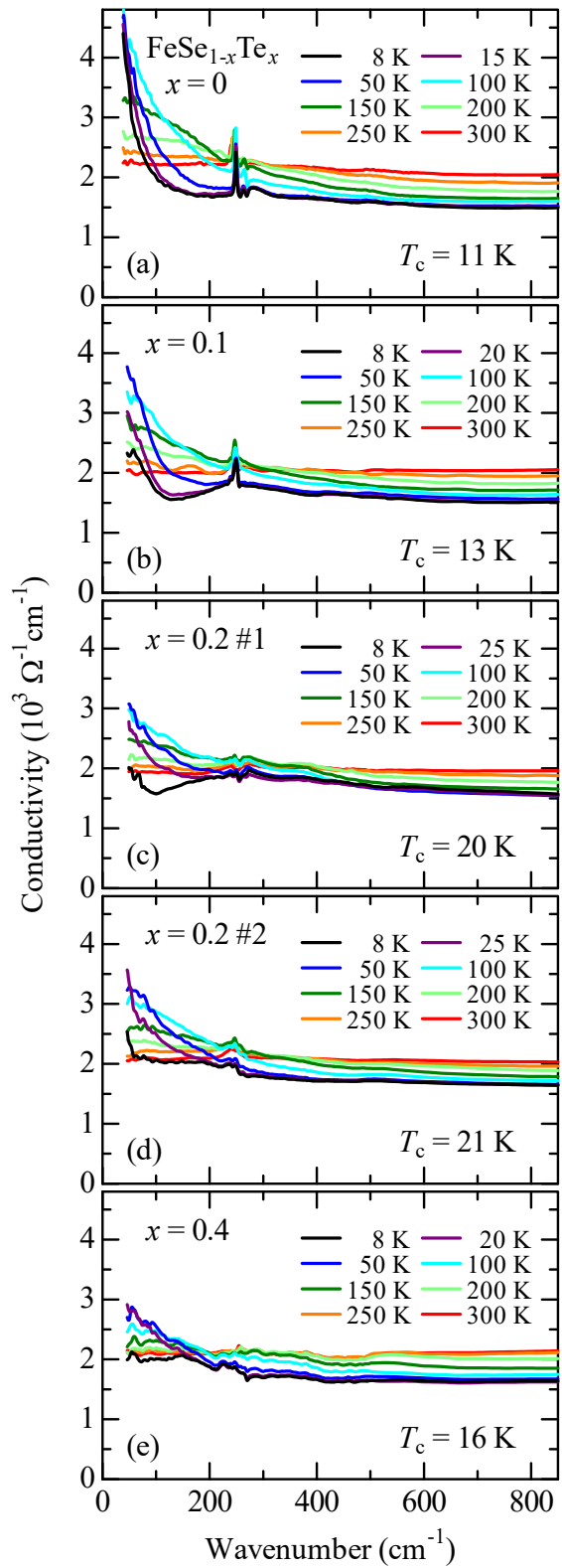}%
\caption{\label{} Temperature dependence of optical conductivity spectra of \FST{} for (a) $x$ = 0, (b) 0.1, (c) 0.2 \#1, (d) 0.2 \#2, and (e) 0.4.}
\end{figure}

Figure 2 shows the temperature dependence of the in-plane optical conductivity spectra for \FST{}. One can see that the substrate contribution has been nicely removed, except for the energy region around 265 \cm{}, corresponding to the optical phonon frequency of substrate \CF{} (see Appendix A). An infrared-active phonon corresponding to an $E_u$ mode, which involves antiphase in-plane motions of Fe and Se/Te atoms, is observed at $\sim$ 249 \cm{} for $x$ = 0 and 0.1. The phonon mode becomes smeared with increasing Te content and is not clearly seen for $x$ = 0.2 and 0.4. For all the compositions shown in Fig.~2, the spectra at high temperatures are almost flat. The absence of an appreciable Drude peak indicates that a coherent component is overdamped, leading to a dominant contribution from incoherent charge dynamics. Such a highly incoherent nature has been commonly observed for optical spectra of iron chalcogenides at high temperatures~\cite{Dai2014,Homes2015}. With decreasing temperature, a Drude peak becomes discernible. This results in an increase in the dc conductivity, consistent with the metallic temperature dependence of resistivity as shown in Figs.~1(b)--(f). The low-temperature Drude peak becomes less and less pronounced with Te substitution. With further decreasing temperature below \Tc{}, a superconducting response is observed as a suppression of optical conductivity, but this is outside the scope of the present study.

\subsection{Drude-Lorentz analysis}

\begin{figure*}
\includegraphics[width=17cm]{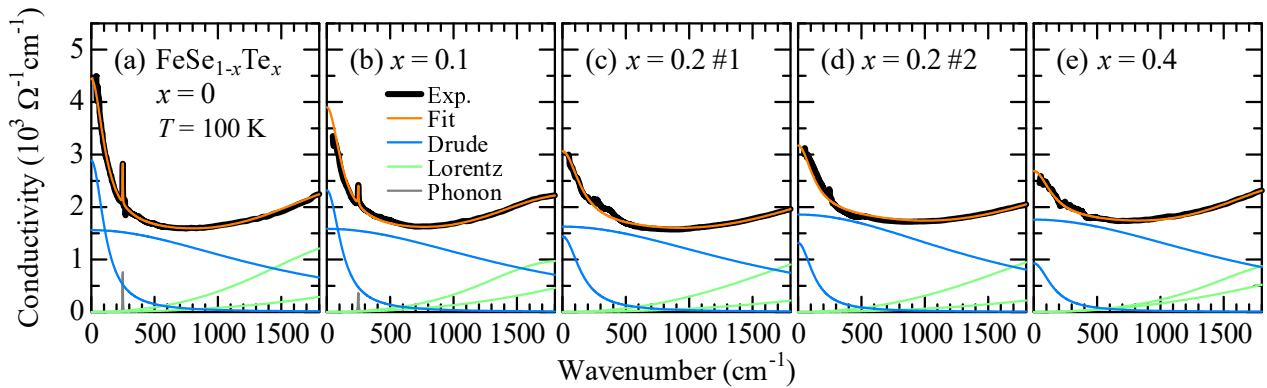}%
\caption{\label{} Decomposition of the optical conductivity spectra of \FST{} at $T$ = 100 K for (a) $x$ = 0, (b) 0.1, (c) 0.2 \#1, (d) 0.2 \#2, and (e) 0.4. For $x$ = 0 and 0.1, the optical phonon mode of \FST{} is clearly seen and is fitted by a Lorentz oscillator.}
\end{figure*}
 
The optical conductivity spectra of \FST{} are characterized by a small Drude peak and an almost flat tail extending up to $\sim$ 1000 \cm{} (see also Appendix B). Thus, we can interpret the low-energy spectrum as a Drude component with small spectral weight present on a component with a weak energy dependence. The latter can be expressed by a very broad Drude term. The presence of multiple Drude components is naturally explained by a multiorbital nature of FeSe. We decomposed the spectrum using a Drude-Lorentz model with a narrow and a broad Drude term. In this model, the complex dielectric function $\tilde{\epsilon}(\omega)$ can be written as
\begin{align*}
\tilde{\epsilon}(\omega) = \; &\epsilon_\infty - \frac{\omega_{\mathrm{p,n}}^2}{\omega^2 + i \omega/\tau_\mathrm{n}} - \frac{\omega_{\mathrm{p,b}}^2}{\omega^2 + i \omega/\tau_\mathrm{b}} \\
&+ \sum_{j} \frac{\Omega_j^2}{\omega_j^2 - \omega^2 - i \gamma_j\omega},
\end{align*}
where $\epsilon_\infty$ is the real part of the dielectric function at high frequency, and $\omega_{\mathrm{p,n}}$ ($\omega_{\mathrm{p,b}}$) and $1/\tau_{\mathrm{n}}$ ($1/\tau_{\mathrm{b}}$) are the plasma frequency and the scattering rate for the narrow (broad) Drude component, respectively. In the last term, $\omega_j$, $\gamma_j$, and $\Omega_j$ are the frequency, width, and strength of the $j$th excitation, respectively. The complex conductivity is expressed by $\tilde{\sigma}(\omega) = \sigma_1 + i \sigma_2 = -i \omega[\tilde{\epsilon}(\omega) - \epsilon_{\infty}]/60$ (in units of $\Omega^{-1}$\cm{}). This model has widely been applied for the analysis of the optical spectra of FeSCs and turned out to well explain the temperature and the composition dependence of the spectrum~\cite{Wu2010,Nakajima2010,Nakajima2014a,Homes2015,Wang2016}.

Figure 3 shows the decomposition of the optical conductivity spectra at $T$ = 100 K. In the low-energy region, the spectrum is dominantly expressed by the narrow and broad Drude terms. The narrow Drude component corresponds to the Drude peak shown in Fig.~2 and is responsible for the temperature dependence. From the width of the Drude term (the scattering rate $1/\tau$) and the band dispersion (the Fermi velocity $v_{\mathrm{F}}$), we can estimate the mean free path $l$ ($= v_{\mathrm{F}}\tau$). While the mean free path for the narrow Drude component is much larger than the lattice constant, that for the broad Drude component is significantly shorter than the shortest interatomic spacing~\cite{Nakajima2014a}. Thus, the charge dynamics represented by the broad Drude component is incoherent. Using the value of the dc conductivity obtained from the present Drude-Lorentz fitting, we calculated the dc resistivity and plotted as a function of temperature in Figs.~1(b)--(f). The temperature dependence is in good agreement with $\rho{}(T)$, indicating that the thin-film fitting analysis gives a reasonable and reliable result.

Given that both electrons and holes contribute to the transport properties in FeSe~\cite{Watson2015b}, the narrow Drude component should arise from electron and hole carriers, although it is difficult to determine which portion of the Fermi surface gives rise to this component. The charge carriers corresponding to the broad Drude component are subject to strong scattering. One of the possible origins is spin fluctuations between hole and electron pockets~\cite{Nakajima2010}. In the present system, the strong electronic correlations, particularly for the $d_{xy}$ orbital, are also a candidate to induce incoherence.

From the decomposed conductivity spectrum (Fig.~3), it is evident that Te substitution leads to a significant reduction of the narrow Drude weight, while the contribution from the broad Drude component shows a slight increase. In addition, the width of the narrow Drude component, corresponding to the scattering rate $1/\tau_{\mathrm{n}}$, increases with increasing $x$, reflecting the disorder effect caused by Te substitution for Se. These give rise to a rapid decrease in the height (the zero-energy value) of the narrow Drude component. As a result, the dc conductivity contributed from the narrow Drude component becomes smaller than that from the broad one for $x$ = 0.2 and 0.4. This suggests that the dc transport properties in \FST{} for higher $x$ are characterized by a largely incoherent nature even at low temperatures.

\begin{figure}
	\includegraphics[width=8cm]{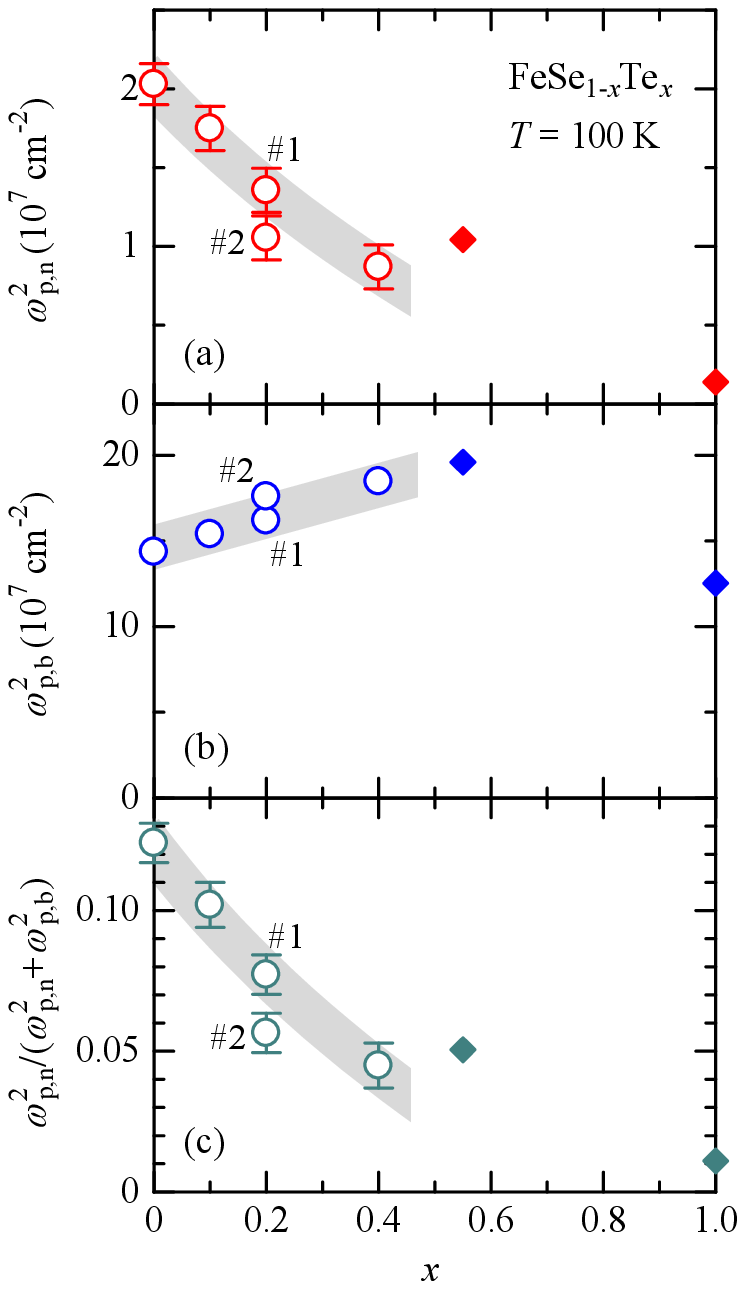}%
	\caption{\label{} Results of the decomposition of the optical conductivity spectra for \FST{} at $T$ = 100 K. The weight of (a) the narrow Drude component \wn{} and (b) the broad Drude component \wb{} is plotted as a function of $x$. (c) Composition dependence of the fraction of the narrow Drude weight \wratio{}. Gray lines are guides to the eye. The data for $x$ = 0.55~\cite{Homes2015} and 1.0~\cite{Dai2014} measured on single crystals are also plotted.}
\end{figure}

The systematic variation of the Drude weights can be quantitatively confirmed in Figs.~4(a) and 4(b), where we plot the weights of the narrow and the broad Drude component (\wn{} and \wb{}, respectively) at 100 K as a function of $x$. The data for single crystals of \FST{} with $x$ = 0.55~\cite{Homes2015} and 1.0~\cite{Dai2014} are also shown. As can be seen in Fig.~3, \wn{} decreases with Te substitution, whereas \wb{} increases. Note that, for $x$ = 0.2, the two samples show the different values for both \wn{} and \wb{}. This indicates that, even though the sample composition is the same, the electronic state varies depending on samples, which likely arises from a difference of a lattice parameter, namely, the degree of in-plane strain. Although \wn{} decreases with Te substitution, the total Drude weight (\wn{} + \wb{}) increases. This is consistent with a recent ARPES study~\cite{Huang2020}, which demonstrated that the size of the Fermi pockets increases with increasing $x$. The composition dependences of \wn{} and \wb{} are roughly followed by the result of the single crystals. \wb{} for FeTe ($x$ = 1.0), however, is much smaller than that even for $x$ = 0. This is probably because the strongly incoherent nature of FeTe gives rise to carrier localization and suppresses the carrier density~\cite{Dai2014}.

\subsection{Evolution of electronic correlations}

The decrease in \wn{} with $x$ can be understood taking into account the evolution of the band structure. Since $\omega_{\mathrm{p,n}}^2 = 4 \pi n / m^*$, where $n$ and $m^*$ stand for the coherent carrier density and effective mass, respectively, the decrease in \wn{} stems from a decrease in $n$ and/or an enhancement of $m^*$. From ARPES measurements, it has been found in \FST{} that a mass enhancement as well as an incoherent electronic state is induced by Te substitution~\cite{Huang2020,Ieki2014,Nakayama2021}. The enhancement of $m^*$ was also observed by the specific-heat study~\cite{Noji2012}. Another insight from the ARPES study is that Te substitution induces the appearance of the $d_{xy}$ orbital character in the hole Fermi surface at the Brillouin-zone center~\cite{Nakayama2021}. This suggests that the coherent region in the momentum space is shrunk due to the mixing of the $d_{xy}$ orbital with a strong mass renormalization, leading to a decrease in $n$ and hence a spectral-weight transfer from \wn{} to \wb{}. Thus, the decrease in \wn{} can be attributed to a combined effect of the enhancement of $m^*$ and the decrease in $n$.

The present results point toward an enhancement of electronic correlations with Te substitution. Optical spectroscopy allows us to estimate the strength of electronic correlation~\cite{Qazilbash2009,Degiorgi2011,Schafgans2012,Nakajima2014b}. A direct way is to calculate the ratio of the experimental kinetic energy, which is associated with the spectral weight of the Drude component, to the theoretically obtained value from the band calculations. The spectral weight contributed by charge carriers can be estimated from the effective carrier number, $N_{\mathrm{eff}}(\omega_{\mathrm{c}}) = \frac{2 m_0 V}{\pi e^2} \int_{0}^{\omega_{\mathrm{c}}} \sigma_1(\omega^\prime) d\omega^\prime$, where $m_0$ and $V$ denote the free electron mass and the cell volume containing one Fe atom, respectively. The cutoff frequency $\omega_{\mathrm{c}}$ should be set to include all of the Drude response but not to include significant contributions from interband transitions. However, it is difficult to unambiguously define $\omega_{\mathrm{c}}$, as interband excitations start at a fairly low energy and overlap with the Drude component. Alternatively, we took the fraction of the coherent spectral weight \wratio{}, or the degree of coherence, as a measure of the strength of electronic correlations~\cite{Nakajima2014b}. This approach gives the estimation solely from the experimental data without any theoretical calculation.

In Fig.~4(c), we plot the fraction of the narrow Drude component \wratio{} as a function of $x$. This fraction severely decreases with increasing $x$, indicative of stronger electronic correlations for higher $x$ in \FST{}. This evolution of the electronic state can be understood in terms of a change in local crystal structure. The substitution of larger Te atoms for smaller Se ones increases the bond length between Fe and chalcogen (\textit{Ch}) atoms and decreases the \textit{Ch}-Fe-\textit{Ch} bond angle~\cite{Imai2017b}. The former makes the overlap between Fe and \textit{Ch} orbitals smaller and hence the band narrower. The latter suppresses the effective hopping of carriers between Fe atoms via \textit{Ch}. These lead to an enhancement of electronic correlations, in line with the increase in $m^*$. Such a behavior was also observed for the case of another isovalent substitution, BaFe$_2$(As$_{1-x}$P$_x$)$_2$~\cite{Nakajima2013,Nakajima2014b}. With substituting smaller P for larger As, the narrow Drude weight systematically increases, indicating a decrease in $m^*$.

The present system exhibits strong orbital-dependent correlation effects, and the $d_{xy}$ orbital plays a crucial role. As mentioned earlier, iron chalcogenides are characterized by strong electronic correlations and the large orbital differentiation, resulting in a large renormalization factor of the $d_{xy}$ orbital~\cite{Yin2011}. In \FST{}, Te substitution gives rise to the enhanced contribution of the $d_{xy}$ orbital to the Fermi-surface construction~\cite{Nakayama2021} and a prominent mass enhancement for the $d_{xy}$ orbital~\cite{Huang2020}. These facts are in good agreement with the present result that the charge dynamics in \FST{} becomes more incoherent with increasing $x$. As the system gets closer to FeTe, the $d_{xy}$ orbital exhibits a Mott-insulating state, while the other orbitals maintain a degree of itinerancy~\cite{Huang2020}. Such a phase, known as an orbital-selective Mott phase, would result in a gap-like feature in the optical spectrum of FeTe~\cite{Dai2014}.

\subsection{Effect of the nematic transition}

\begin{figure}
\includegraphics[width=8.3cm]{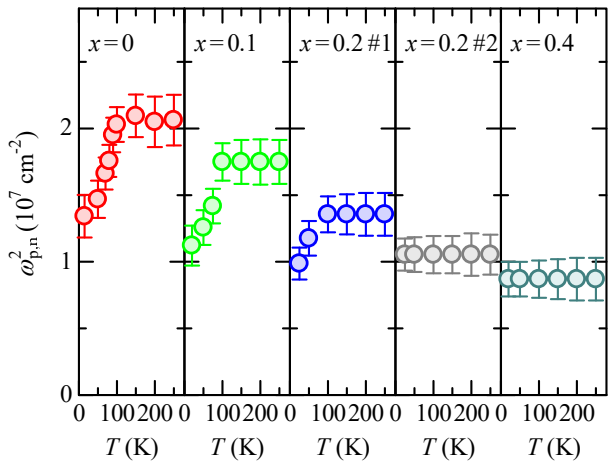}%
\caption{\label{} Temperature dependence of \wn{} for \FST{} ($x$ = 0, 0.1, 0.2 \#1, 0.2 \#2, and 0.4). The reduction of \wn{} at low temperatures indicates the presence of the structural phase transition.}
\end{figure}

The nematic transition has an impact selectively on the narrow Drude component. In Fig.~5, we show the temperature dependence of \wn{} for \FST{}. In the high-temperature tetragonal phase, \wn{} does not change with temperature and exhibits a systematic decrease with $x$ as already shown in Fig.~4(a). As demonstrated in our previous study~\cite{Nakajima2017}, \wn{} for $x$ = 0 shows a decrease upon entering the nematic phase below \Ts{}. We attributed this to the decrease in the coherent carrier density arising from gradual modification of the Fermi surface due to the development of the orbital order. The reduction of the carrier density below \Ts{} is also observed by terahertz magneto-optical spectroscopy on an FeSe thin film on LaAlO$_3$ substrate~\cite{Yoshikawa2019}. Note that \wb{} remains unchanged or shows a weak increase in the nematic phase~\cite{Nakajima2017}. This is in stark contrast with the case for iron arsenides showing a clear gap-like feature in the antiferromagnetic orthorhombic phase~\cite{Hu2008,Nakajima2010,Charnukha2013,Dai2016}. The magnetostructural transition severely affects the broad Drude component~\cite{Nakajima2010,Dai2016}, suggesting that spin fluctuations play a dominant role to produce strong carrier scattering.

As in the case for $x$ = 0, a decrease in \wn{} was also observed for $x$ = 0.1 and 0.2 \#1, whereas a suppression of \wn{} was absent for $x$ = 0.2 \#2 and 0.4. Since the decrease in \wn{} manifests the presence of the nematic phase, the present result evidences that the nematic transition is completely suppressed at $x \sim$ 0.2, consistent with the transport~\cite{Imai2017} and ARPES~\cite{Nakayama2021} studies. For $x$ = 0, 0.1, and 0.2 \#1, compared with the high-temperature value above \Ts{}, \wn{} decreases by around one third at the lowest temperature. This large reduction would stem from an effective Fermi energy comparable with the energy splitting of $d_{xz}$ and $d_{yz}$ bands due to the orbital ordering. Interestingly, the two samples for $x$ = 0.2 showed the distinct behaviors. This indicates that the end point of the nematic transition for \FST{} thin films on \CF{} substrates is around $x$ = 0.2, which can be altered depending on the degree of in-plane strain. Indeed, it was reported that the Te content necessary to suppress the nematic transition can be controlled by changing the substrate materials~\cite{Imai2017}. There seems to be a threshold of the high-temperature value of \wn{}. The nematic transition would be present for the samples with \wn{} larger than $\sim 1.1 \times 10^7$ cm$^{-2}$.

The reduction of the coherent carrier density induced by the nematic transition should affect superconductivity. In the present system, the carrier density seems to correlate with \Tc{}~\cite{Nabeshima2018,Nabeshima2020,Nakajima2021}. In this context, the presence of the nematic transition, which reduces the coherent carrier density, hinders superconductivity, although it is not clear whether the presence of the nematic order itself competes with superconductivity. For $x$ = 0.2 \#1, we confirmed the nematic transition as well as the two-step superconducting transition [inset of Fig.~1(d)], suggesting that the sample contains two parts: one showing the nematic transition with lower \Tc{} and the other showing no nematic transition with higher \Tc{}. This infers that a slight difference of in-plane strain results in elimination of the nematic transition and a large enhancement of \Tc{}. The transport study on \FST{} single crystals demonstrated that suppression of the nematic phase under pressure gives rise to an enhancement of \Tc{}~\cite{Mukasa2021}, which may be related with the increase in the carrier density. Unlike \FST{}, \Tc{} for S-substituted FeSe shows no drastic change or rather a decrease across the boundary of the nematic phase~\cite{Nabeshima2018b,Yi2020,Reiss2017}. In this case, no significant change of the carrier density across the phase boundary~\cite{Nabeshima2020}. The origin of the contrasting behavior in \FSS{} should be clarified by further studies.

Superconductivity should also be influenced by the change in the correlation strength. Outside the nematic phase, \Tc{} decreases with increasing $x$. Probably, electronic correlations for higher values of $x$ are too strong to support high-\Tc{} superconductivity. The value of \wratio{} for $x$ = 0.2 with maximal \Tc{} is $\sim$ 0.07, which is close to that for doped BaFe$_2$As$_2$~\cite{Nakajima2014b}. Thus, the strength of electronic correlations is commonly an important ingredient not only for iron pnictides but also for iron chalcogenides. In light of electronic correlations, \FST{} films on \CF{} substrates for $x < 0.2$ have a potential to exhibit superconductivity with \Tc{} comparable with or higher than that for $x$ = 0.2, but actual \Tc{} is much lower because of the suppression of the coherent carrier density due to the nematic order.

Finally, we comment on the influence of spin fluctuations. In the temperature-pressure phase diagram of FeSe, \Tc{} takes a maximum at the boundary of the magnetic phase~\cite{Sun2016}. The NMR study observed enhanced spin fluctuations for \FSS{} in the nematic phase near the composition with maximal \Tc{}~\cite{Wiecki2018}. These results indicate the intimate relationship between spin fluctuations and superconductivity. Although no magnetic phase has been observed for \FST{} both for the composition and the pressure axis~\cite{Imai2017,Mukasa2021}, the appearance of the $d_{xy}$ orbital in the hole Fermi surface induced by Te substitution can enhance spin fluctuations in the $d_{xy}$ channel~\cite{Nakayama2021}. The enhancement of \Tc{} may partly result from the presence of spin fluctuations associated with the nesting between $d_{xy}$ portions of the hole and electron Fermi pockets.

\begin{figure}
	\includegraphics[width=7.8cm]{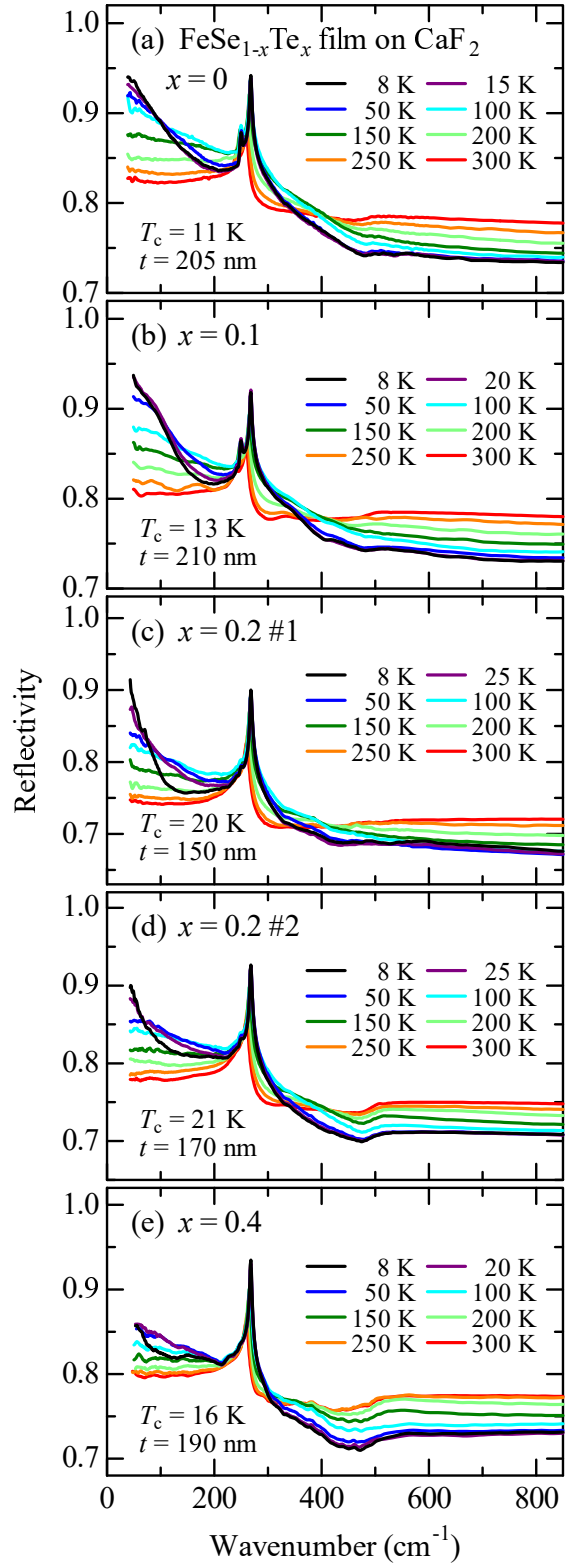}%
	\caption{\label{} Temperature dependence of the reflectivity spectrum in the low-energy region for the \FST{} thin films on \CF{} substrates. $t$ is a thickness of the films. The spectrum is largely affected by substrate \CF{}, as described in the text.}
\end{figure}

\begin{figure*}
	\includegraphics[width=17cm]{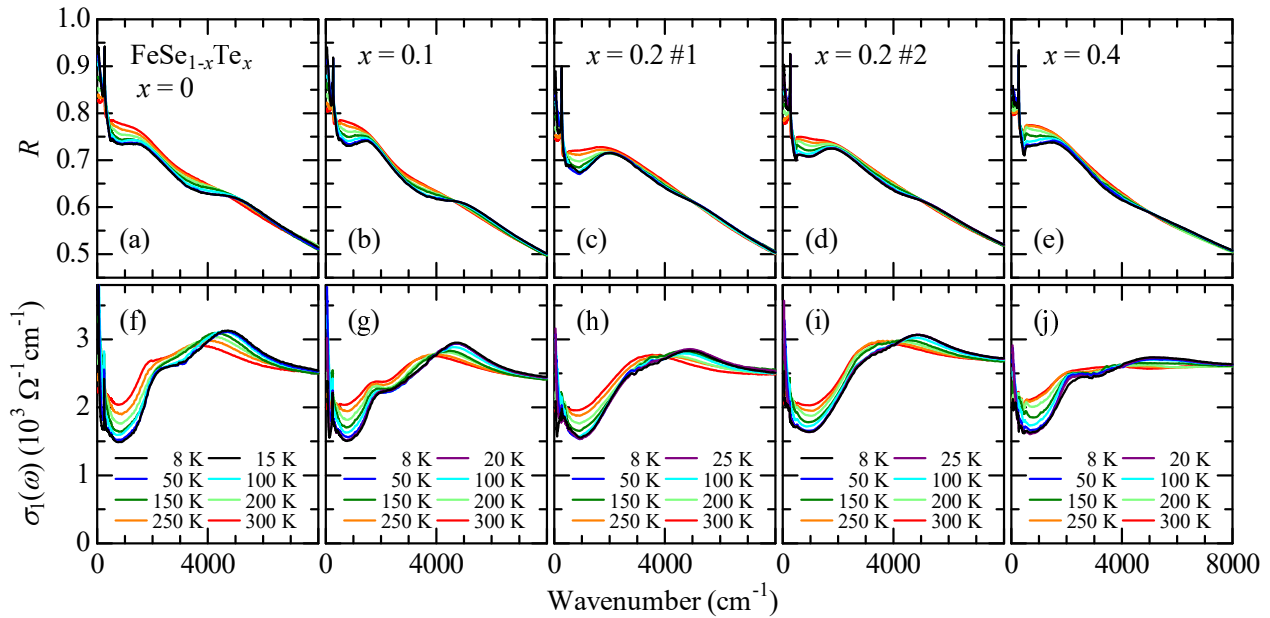}%
	\caption{\label{} Temperature dependence of (a)--(e) the reflectivity spectrum for the \FST{} thin films and (f)--(j) the optical conductivity spectrum for \FST{} in a wide energy range up to 8000 \cm{}.}
\end{figure*}

\section{Conclusion}
 
We measured in-plane optical spectrum of the \FST{} thin films on \CF{} substrates for $x$ = 0, 0.1, 0.2, and 0.4. For all the compositions, the low-energy optical conductivity spectra at high temperatures are almost flat, indicative of the highly incoherent charge dynamics. With decreasing temperature, the Drude response becomes appreciable, the degree of which is weakened with Te substitution. The Drude-Lorentz analysis revealed that \wn{} decreases with increasing $x$, while \wb{} increases. The reduction of the fraction of \wn{} is attributed to an enhancement of electronic correlations with Te substitution due to the change in the local crystal structure. The $d_{xy}$ orbital is considered to play a crucial role in the evolution of electronic correlations. Upon entering the low-temperature nematic phase, \wn{} decreases. This behavior disappeared at $x \sim$ 0.2, corresponding to the end point of the nematic transition. The suppression of \wn{} was observed for one of the two samples with $x$ = 0.2 but not for the other, which is consistent with the fact that \Tc{} abruptly jumps up when suppressing the nematic phase. The present result suggests that the enhancement of \Tc{} originates from the increase in the coherent carrier density due to the suppression of the nematic transition. The decrease in \Tc{} with Te substitution for $x > 0.2$ likely arises from too strong electronic correlations, which is harmful to high-\Tc{} superconductivity.

\begin{acknowledgments}
This work was supported by JSPS KAKENHI Grant Numbers JP18K13500 and JP18K03513.
\end{acknowledgments}

\appendix
\section{Low-energy reflectivity spectra}

Figure 6 shows the temperature dependence of optical reflectivity spectra of \FST{} films on \CF{} substrates in the low-energy region. The reflection from the interface between \FST{} and \CF{} strongly affects the spectra. One can see a tendency that the reflectivity is lower for thinner films. The peak at $\sim$ 265 \cm{} corresponds to the optical phonon mode of substrate \CF{}. For $x$ = 0 and 0.1, the optical phonon mode of FeSe is discernible at $\sim$ 249 \cm{} on the left shoulder of the phonon mode of \CF{}. The reflectivity spectra at high temperatures are almost flat below 200 \cm{}. This feature arises from \FST{} with a highly incoherent nature grown on insulating \CF{}. With decreasing temperature, the reflectivity at lowest-energy region increases, in agreement with a metallic behavior of \FST{}.

\section{Optical spectra for a wide energy range}

Figure 7 shows the temperature dependence of reflectivity spectra of \FST{} films and extracted optical conductivity spectra of \FST{} for a wide energy range up to 8000 \cm{}. We observed no appreciable temperature dependence above 8000 \cm{}. In the conductivity spectra, two peaks are present at $\sim$ 2000 \cm{} and $\sim$ 4000 \cm{}, corresponding to interband transitions [Figs.~7(f)--(j)]. Note that the two peaks become less clear with increasing $x$. This tendency is more pronounced for the peak at $\sim$ 4000 \cm{}. For $x$ = 0.4 at 300 K, only a gentle hump structure can be seen at $\sim$ 3500 \cm. With decreasing temperature, the two peaks sharpen and show a blueshift or spectral weight transfer to higher energies. This behavior, which is seen in various FeSCs~\cite{Wang2012}, is especially prominent for the higher-energy interband transition. The value of the optical conductivity at $\sim$ 4000 \cm{} does not vary with temperature, indicative of an isosbestic point at the corresponding energy. The double-peak structure is observed for all the compositions investigated in the present study, but the 4000-\cm{} peak significantly broadens and becomes ambiguous with increasing $x$. This is consistent with the optical spectrum for FeTe ($x$ = 1.0)~\cite{Dai2014,Wang2012}, in which the higher-energy peak cannot be recognized.


\providecommand{\noopsort}[1]{}\providecommand{\singleletter}[1]{#1}%

\end{document}